\documentclass{ws-procs961x669}
\begin{document}
\title{Linear perturbations in Horndeski theories with spatial curvature}
\author{Serena Gambino}
\address{Scuola Superiore Meridionale (SSM), Largo San Marcellino 10, 80138 Napoli, Italy\\
Istituto Nazionale di Fisica Nucleare (INFN), Sezione di Napoli Complesso
Universitario Monte S. Angelo, Via Cinthia 9 Edificio G, 80138 Napoli, Italy.\\
\email{s.gambino@ssmeridionale.it}}
\author{Francesco Pace}
\address{Dipartimento di Fisica, Universit\`a degli Studi di Torino, Via P. Giuria 1, 10125, Torino, Italy\\
Istituto Nazionale di Fisica Nucleare (INFN), Sezione di Torino, Via P. Giuria 1, 10125, Torino, Italy\\
Istituto Nazionale di Astrofisica (INAF), Osservatorio Astrofisico di Torino, 10125, Pino Torinese, Italy.\\
\email{francesco.pace@unito.it}}
\begin{abstract}
We analyse the implications of the presence of spatial curvature in modified gravity models. As it is well known, the current standard cosmological model, the $\Lambda$CDM, is assumed to be spatially flat based on the results of many experiments. However, this statement does not necessarily hold for a modified gravity (MG) model, and this leads to couplings of the spatial curvature with the parameters of the chosen cosmological model. In this paper, we illustrate the theoretical development of how spatial curvature affects the equations of motion at linear order for scalar and tensor perturbations modes using a model-independent approach based on the formalism of the Effective Field Theory (EFT) of dark energy (DE).\\
The results show that spatial curvature gives rise to a coupling with the scalar field perturbations and the functions parameterizing the model.
\end{abstract}

\bodymatter

\section{Introduction}\label{intro}
The $\Lambda$CDM model has been remarkably successful in describing the accelerated expansion of the universe, attributing it to a cosmological constant ($\Lambda$) and the formation of structure by the presence of cold dark matter (CDM). We point out that one of the results of the model is spatial flatness (spatial curvature $\mathcal{K}$ extremely close to zero), confirmed by the results obtained from the Planck satellite data \cite{PLANCK2018}.\\
Despite the success of the Standard Model in describing our universe, it still faces several major challenges. Among these, we recall the Hubble and $\sigma_8$ tensions, two well-known problems based on the incompatibility of the values of these cosmological parameters measured in two different epochs. These, among other problems, have led over the years to questions about whether the $\Lambda$CDM is really the right model to describe our universe. \\
To address and try to find an answer to these issues, we examine a modified gravity (MG) model called Horndeski model, focusing on the effects of a non-zero spatial curvature. Indeed, setting $\mathcal{K}$ to zero has been assumed in many modified gravity models, as an extrapolation of the results of the $\Lambda$CDM, and this may overlook important effects on perturbations and parameter constraints. In particular, we extend the previous analysis of both scalar and tensor linear perturbations of Horndeski models to space-times endowed with spatial curvature using the formalism developed by the EFT of DE \cite{GLEYZES}. By incorporating spatial curvature, we aim to assess, with future numerical analysis, whether its inclusion can alleviate current cosmological tensions and provide a more accurate framework for the evolution of the universe. 

\section{Theoretical framework and results}
\label{theory}
\subsection{EFT of dark energy - Cosmological background}
The EFT of DE is a framework that unifies the class of single-scalar field DE/MG cosmological models in a model-independent way to study cosmic acceleration using a perturbative approach. It introduces additional degrees of freedom beyond the standard Friedmann cosmological background.
This formalism allows for a simplified confrontation of different models with the data. For a detailed explanation, see also \cite{FRUSCIANTE}.\\
To construct the theory, the unitary gauge is chosen, where the perturbations of the additional scalar degrees of freedom (DoF) are absorbed into the gravitational sector. The scalar perturbations are encoded by the metric components, and the action is organised with geometrical operators compatible with the residual symmetries. The system of natural units, $c=\hbar=G=1$, will be used throughout the discussion. The EFT action at the background level can be written in a general form as
\begin{equation}
S = \int d^4x \sqrt{-g} \left[ \frac{M_{*}^2}{2} f(t) R - \Lambda(t) - c(t) g^{00} \right] + S_m[g_{\mu\nu}, \chi_{\rm m}]\,,
\label{bck_action}
\end{equation}
where $M_{*}$ is the bare Planck mass, $R$ is the Ricci scalar, and the functions $f(t)$, $\Lambda(t)$, and $c(t)$ describe deviations from GR. The matter fields $\chi_{\rm m}$ are minimally coupled to the metric $g_{\mu\nu}$. 
As anticipated, the background is typically described by an isotropic and homogeneous metric, the Friedmann-Lemaître-Robertson-Walker (FLRW) metric. We choose the following form
\begin{equation}
    ds^2=-dt^2+a^2(t) g_{ij}dx^idx^j, \qquad \gamma_{ij}=\frac{\delta_{ij}}{(1+\frac{\mathcal{K} r^2}{4})^2}\,,
\end{equation}
here $a(t)$ is the scale factor and $\mathcal{K}$ the spatial curvature. The background equation of motion obtained from Eq.~(\ref{bck_action}) are 
\begin{align}
    &\, 3 M_{*}^{2}f\left[H^2+H\dot{f}/f+\frac{\mathcal{K}}{a^{2}}\right]-\rho_{m} = -(\Lambda-c)\,,\\
    &\, M_{*}^{2}f\left[3H^2+2\dot{H}+\frac{\mathcal{K}}{a^{2}}+2H\dot{f}/f+\ddot{f}/f\right]+p_{m} = \Lambda-c \,.
\end{align}
\subsection{EFT of dark energy - Linear perturbation level}
We will focus now on the linear order of perturbations. The EFT framework allows any scalar-tensor theory to be mapped into its language by translating the parameters and fields of the theory into the corresponding EFT functions and operators. This is achieved by a procedure involving the Arnowitt-Deser-Misner (ADM) decomposition \cite{GLEYZES}. This formalism is introduced as a tool to rewrite a general scalar-tensor Lagrangian and the EFT action, and it allows the description of gravitational theories containing higher-order spatial derivatives. 
The final form of the Lagrangian is then parameterized with the so-called phenomenological $\alpha$ basis, where time-dependent functions $\alpha_{\rm i}$ characterize the deviations from GR. One can identify all the pieces in the EFT action that correspond to the pieces in the ADM action and connect the theory. For this purpose we used a Horndeski model, and the final expression of the action up to the quadratic order, with the EFT formalism \cite{GLEYZES, GUBITOSI}, is 
\begin{small}
\begin{align}
   S^{(2)} = \int d^4 x \sqrt{-g} \Bigg[ &\, \frac{M_2^4(t)}{2} (\delta g^{00})^2 
    - \frac{m_3^3(t)}{2} \delta K \delta g^{00}  
    - m_4^2(t) \left(\delta K^2 - \delta K^i{}_j \delta K^j{}_i\right) \nonumber \\
    &\, \left. - \frac{\tilde{m}_4^2(t)}{2} \, \delta^{(3)}\!R \, \delta g^{00} \right]\,,
    \label{EFT_action}
\end{align}
\end{small}
where the $M^{\beta}_{\alpha}(t)$ and $m^{\beta}_{\alpha}(t)$ are the EFT functions, $K, K^j{}_i$ the extrinsic curvature scalar and tensor and $\delta^{(3)}\!R, R$ respectively the three- and four-dimensional Ricci scalar.\\
To obtain also an explicit equation for the evolution of the scalar field, we need to make it explicit in the action by restoring full diffeomorphism invariance using the {\it St\"uckelberg trick}; by so, we have to force back the broken gauge transformation on the field in the Lagrangian by imposing a time coordinate transformation: $ t \rightarrow \tilde{t}= t + \pi(x^\mu)$ where $\pi$ is the perturbation of the extra DoF.

We show now the final results obtained and described in the Newtonian gauge
\begin{equation}
    ds^2 = -(1+2\Phi)\mathrm{d}t^2 + a^{2}(1-2\Psi)\gamma_{ij} dx^{i}dx^{j}\,,
\end{equation}
with the following perturbed stress-energy tensor
\begin{equation}
    \delta T^{0} _{0}=-\delta\rho_{m},\quad  \delta T^{0} _{i}=\partial_i q_m=(\rho_{m}+p_{m})\partial_i v_m, \quad \delta T^{i} _{j}=\delta p_{m}\delta^{i}_{j}+(D^iD_j-\frac{1}{3}\delta^i_jD^2)\sigma_m\,,
\end{equation}
where $D^2$ identifies the covariant spatial second order derivative. We obtain, fixing the $\alpha$-basis, for the (00) component of the field equations
\begin{small}
\begin{align}  
   &\,  6(1+\alpha_{\rm B})H\dot{\Psi}+(6-\alpha_{\rm K}+12\alpha_{\rm B})H^2\Phi-2(1+\alpha_{\rm H})\frac{(D^2+3 \mathcal{K})}{a^2}\Psi  +(\alpha_{\rm K}-6\alpha_{\rm B})H^2\dot{\pi} \nonumber \\
   &\, +6H\left[\frac{\rho_m+p_m}{2M^2}+\dot{H}(1+\alpha_{\rm B})+\frac{\mathcal{K}}{a^2}(1-\alpha_{\rm B}) -\frac{1}{3}(\alpha_{\rm H}-\alpha_{\rm B})\frac{(D^2+3 \mathcal{K})}{a^2}\right]\pi=-\frac{\delta\rho_m}{M^2} \,,
   \label{00}
\end{align}
\end{small}
then, for the (0i) component:
\begin{equation}
   2\dot{\Psi}+2(1+\alpha_{\rm B})H\Phi-2H\alpha_{\rm B}\dot{\pi}+\left(2\dot{H}+\frac{\rho_{m}+p_{m}}{M^2}\right)\pi=-\frac{(\rho_{m}+p_{m})v_{m}}{M^2} \,,
\label{0i}
\end{equation}
and the trace of the (ij) components:
\begin{small}
\begin{align}  
    &\, 2\ddot{\Psi}+2(3+\alpha_{\rm M})H\dot{\Psi}+2(1+\alpha_{\rm B})H\dot{\Phi} +2\left[\dot{H}+\frac{\rho_m+p_m}{2M^2}-(H\alpha_{\rm B})\dot{}-(3+\alpha_{\rm M})\alpha_{\rm B} H^2\right]\dot{\pi} \nonumber\\
  &\, +2\left[\dot{H}-\frac{\rho_m+p_m}{2M^2}+(H\alpha_{\rm B})\dot{} +(3+\alpha_{\rm M})(1+\alpha_{\rm B})H^2\right]\Phi+2\left[(3+\alpha_{\rm M})H\dot{H}+\frac{\dot{p}_m}{2M^2} \right. \nonumber\\
   &\, \left. +\ddot{H}\right]\pi  -2H\alpha_{\rm B}\ddot{\pi} -6\frac{{{ \mathcal{K}}}}{a^2}[\dot{\pi}+(1+\alpha_{\rm M})H\pi]
=\frac{1}{M^2}\left(\delta p_m + \frac{2}{3}\frac{(D^2+3\mathcal{K})}{a^2}\sigma_m\right)\,.
   \label{ijtr}
      \end{align}
      \end{small}
The (ij)-traceless components:
\begin{small}
\begin{equation}
     (1+\alpha_{\rm H})\Phi - (1+\alpha_{\rm T})\Psi +(\alpha_{\rm M}-\alpha_{\rm T})H\pi-\alpha_{\rm H}\dot{\pi}=-\frac{\sigma_m}{M^2}     \,,
\label{ijtrless}
\end{equation}
\end{small}
and lastly, the equation of motion for $\pi$:
\begin{small}
\begin{align} 
  &\, H^2\alpha_{\rm K}\ddot{\pi} + 6H\alpha_{\rm B}\ddot{\Psi} + H^2(6\alpha_{\rm B}-\alpha_{\rm K})\dot{\Phi}+\left[\left(H^2(3+\alpha_{\rm M})+\dot{H}\right)\alpha_{\rm K} + (H\alpha_{\rm K})\dot{} \ \right]H\dot{\pi}   \nonumber \\
  &\, +2\frac{(D^2+3{{ \mathcal{K}}})}{a^2}\left[\dot{H}+\frac{\rho_m+p_m}{2M^2}+H^2[1+\alpha_{\rm B}(1+\alpha_{\rm M})+\alpha_{\rm T}-(1+\alpha_{\rm H})(1+\alpha_{\rm M})]   \right.\nonumber \\ 
  &\, \left. +(H(\alpha_{\rm B}-\alpha_{\rm H}))\dot{} \ \right]\pi +6\left[\dot{H}\left(\dot{H}+\frac{\rho_m+p_m}{2M^2}\right)+\dot{H}\alpha_{\rm B}[H^2(3+\alpha_{\rm M})+\dot{H}]+H(\dot{H}\alpha_{\rm B})\dot{} \ \right]\pi  \nonumber\\
  &\, +6\left[\dot{H}+\frac{\rho_m+p_m}{2M^2}+ H^2(3+\alpha_{\rm M})\alpha_{\rm B}+(\alpha_{\rm B} H)\dot{}\right]\dot{\Psi} -\frac{6{\mathcal K}}{a^2}\left[H^2\alpha_{\rm B}(1-\alpha_{\rm M})+(H\alpha_{\rm B})\dot{}\ \right]\pi\nonumber\\
  &\, +\left[6\left(\dot{H}+\frac{\rho_m+p_m}{2M^2}\right)+H^2(6\alpha_{\rm B}-\alpha_{\rm K})(3+\alpha_{\rm M}) +2(9\alpha_{\rm B}-\alpha_{\rm K})\dot{H} +H(6\dot{\alpha_{\rm B}}-\dot{\alpha_{\rm K}})\right] H\Phi\nonumber\\
  &\, +2\frac{(D^2+3{{ \mathcal{K}}})}{a^2}\left[\alpha_{\rm H}\dot{\Psi}+ [H(\alpha_{\rm M} + \alpha_{\rm H}(1+\alpha_{\rm M})-\alpha_{\rm T}) +\dot{\alpha}_{\rm H}]\Psi + H(\alpha_{\rm H}-\alpha_{\rm B})\Phi\right]=0 \,.
  \label{pi}
\end{align} 
\end{small}
Instead, for the tensor part of the linear perturbations, we have the following metric 
\begin{small}
\begin{equation}
    ds^2 = -dt^2 + g_{ij} dx^{i}dx^{j},\ g_{ij}=a^{2}(t)\left(\gamma_{ij} +h_{ij}(t,\vec{x})\right)\,,
\end{equation}
\end{small}
and using the action in Eq.~(\ref{EFT_action}), we obtain the following perturbed field equations
\begin{small}
 \begin{equation}
 \ddot{h}_{ij} +( 3+\alpha_{\rm M})H\dot{h}_{ij} + \frac{1}{a^{2}}(1+\alpha_{\rm T}){{(2\mathcal{K} - D^2)}}h_{ij} = \frac{2}{M^{2}}\left(T_{ij} - \frac{1}{3}T\gamma_{ij}\right)^{TT}\,,
\end{equation}
\end{small}
where $TT$ identify the traceless and transverse part of the stress-energy tensor $T^{\mu\nu}$.
\section{Discussion}
The main goal of this work was to see the effect of spatial curvature in the derived equations and its interplay with the $\alpha$ functions and the perturbations of the scalar field $\pi$ and its derivatives. This might show, with the implementation of the previously derived equations in an Einstein-Boltzmann code, that spatial curvature may modify current constraints on these functions. Understanding whether cosmological tensions are indeed affectedby this coupling and can be alleviated is the final test that this work aims to achieve.

\vfill
\pagebreak

\end{document}